\documentclass[final,5p,times,twocolumn]{elsarticle}

\usepackage{amsmath,amssymb,amsfonts}
\usepackage{graphicx}
\usepackage{textcomp}
\usepackage{lscape}
\usepackage{float}
\usepackage{xcolor}
\usepackage{orcidlink}
\usepackage{longtable}

\usepackage{booktabs}
\usepackage{rotating}
\usepackage{listings}
\usepackage{algpseudocode}
\usepackage{booktabs}
\usepackage[linesnumbered,ruled,vlined]{algorithm2e} 
\usepackage{hyperref}
\def\BibTeX{{\rm B\kern-.05em{\sc i\kern-.025em b}\kern-.08em
    T\kern-.1667em\lower.7ex\hbox{E}\kern-.125emX}}

\usepackage{tabularx,booktabs}
\usepackage[table]{xcolor}
\usepackage{threeparttable}
\usepackage{tikz}
\usepackage{array} 
\usepackage{multirow}

\newcolumntype{L}[1]{>{\raggedright\arraybackslash}p{#1}}
\definecolor{grey1}{HTML}{f5f5f5}

\newcolumntype{C}{>{\centering\arraybackslash}X} 
\journal{Journal}

\begin{document}

\begin{frontmatter}



\title{CAFE-GB: Scalable and Stable Feature Selection for Malware Detection via Chunk-wise Aggregated Gradient Boosting}


\author[label1]{Ajvad Haneef K \corref{cor1} \orcidlink{0000-0002-2613-7131}}
 \cortext[cor1]{Corresponding author}
\ead{ajvad_p210054cs@nitc.ac.in}
\author[label1]{Karan Kuwar Singh}
\ead{karan_m220284cs@nitc.ac.in}
\author[label1]{Madhu Kumar S D\orcidlink{0000-0002-5276-8842}}
\ead{madhu@nitc.ac.in}

\affiliation[label1]{organization={Department of Computer Science and Engineering},
            addressline={National Institute of Technology Calicut}, 
            country={India}}

\begin{abstract}
High-dimensional malware datasets often exhibit severe feature redundancy, instability, and scalability limitations, which undermine the effectiveness and interpretability of machine learning–based detection systems. Although feature selection is widely adopted to address these challenges, most existing methods rely on global importance estimation and lack stability guarantees when applied to large-scale and heterogeneous malware data. To address this gap, this paper proposes CAFE-GB (Chunk-wise Aggregated Feature Estimation using Gradient Boosting), a scalable feature selection framework that derives stable feature rankings by aggregating local importance estimates computed over overlapping data chunks. Unlike conventional approaches that estimate feature relevance globally, CAFE-GB captures locality-aware importance patterns and aggregates them into a globally consistent ranking, improving robustness without increasing model complexity. Feature budget selection is performed separately through a systematic k-selection and stability analysis, enabling principled dimensionality reduction. The proposed framework is evaluated on two large-scale malware datasets, BODMAS and CIC-AndMal2020, using multiple classifiers, statistical significance testing, redundancy analysis, SHAP-based explainability, and runtime and memory profiling. Experimental results show that CAFE-GB reduces feature dimensionality by over 95\% while achieving performance parity with full-feature baselines across Accuracy, F1-score, MCC, ROC-AUC, and PR-AUC. Paired Wilcoxon tests confirm that no statistically significant performance degradation is introduced. Additional analyses demonstrate low feature redundancy, improved interpretability, and reduced downstream computational overhead. Overall, CAFE-GB provides a stable, interpretable, and scalable alternative to conventional feature selection methods for large-scale malware detection.
\end{abstract}




\end{frontmatter}


\section{Introduction}
\label{sec:introduction}
The rapid growth and increasing sophistication of malware have driven widespread adoption of machine learning–based detection systems capable of analyzing large volumes of executable and behavioral data\cite{gorment2023machine,gibert2020rise}. Modern malware datasets are inherently high-dimensional, comprising thousands of static and dynamic features extracted from heterogeneous sources\cite{jeon2024static}. While such rich representations offer strong discriminative potential, they also introduce feature redundancy, instability, and scalability challenges that directly affect model generalization, interpretability, and computational efficiency\cite{gorment2023machine,aboaoja2022malwareissues}. Most existing malware detection research prioritizes improving predictive performance through complex learning architectures, ensemble methods, or deep neural networks\cite{xiong2024modified,vasan2025advanced}. Although these approaches often achieve high accuracy, they exacerbate scalability and interpretability concerns and provide limited insight into the stability of the underlying feature representations. In contrast, comparatively few studies systematically investigate feature selection mechanisms that remain stable, reproducible, and scalable when applied to large-scale malware datasets. Conventional feature selection methods typically rely on global importance estimates computed over entire datasets, making them sensitive to class imbalance, local distributional variations, and sampling bias \cite{theng2024feature,moslemi2023tutorial}. As a result, existing approaches often produce unstable and redundant feature subsets that vary across data partitions, limiting their practical reliability.  The malware datasets typically contain of a large collection of features, necessitating the identification of relevant attributes to optimize model training and testing. In addition the size of the dataset is very large, which demands substantial memory and powerful hardware resources. The contemporary feature selection techniques are not efficient in handling such large-scale, high-dimentional datasets. Most of the feature selection techniques work on the entire dataset, which is entails high computational cost and substantial memory resources.

Among the many challenges in malware detection—such as adversarial manipulation, concept drift, and real-time constraints—this work focuses specifically on the problem of stable and scalable feature selection for high-dimensional malware datasets. Addressing feature selection at this level provides a principled means of improving efficiency and interpretability without increasing model complexity. To this end, we propose CAFE-GB (Chunk-wise Aggregated Feature Estimation using Gradient Boosting), a scalable feature selection framework designed to produce stable, compact, and non-redundant feature subsets for malware detection. We further conduct a systematic k-selection and stability analysis to identify a fixed feature budget that balances robustness and efficiency. The proposed framework is evaluated on two large-scale malware datasets—BODMAS\cite{yand2021bodmas} and CIC-AndMal2020\cite{rahali2020didroid}—using multiple classifiers, statistical significance testing, redundancy analysis, SHAP-based explainability, and runtime and memory profiling. The results demonstrate that CAFE-GB preserves detection performance while substantially reducing feature dimensionality, offering a stable and scalable alternative to full-dimensional malware detection models.

The key contributions in this paper are summarized as follows:
\begin{itemize}
   \item We propose CAFE-GB, a chunk-wise feature selection framework that aggregates local gradient boosting importance estimates to obtain stable global feature rankings for high-dimensional malware data.

    \item We show that chunk-wise aggregation improves feature selection robustness and reproducibility, addressing stability issues overlooked in existing malware detection studies.

    \item We demonstrate that CAFE-GB produces compact, low-redundancy feature subsets while preserving detection performance under aggressive dimensionality reduction.

    \item Extensive experiments confirm that CAFE-GB achieves performance parity with full-feature baselines while reducing feature dimensionality by over 95
\end{itemize}

\section{Related Works}
\label{sec:related-works}

Malware detection has been widely studied using machine learning techniques, with prior research focusing on feature engineering, feature selection, classification models, and explainability \cite{deldar2023deep,daniel2023optimal,saqib2024comprehensive}. This section reviews key advances in these areas and highlights the limitations that motivate the proposed CAFE-GB framework.

\subsection{Machine Learning-Based Malware Detection}

Early malware detection systems relied on signature-based techniques, which are ineffective against obfuscated and evolving threats \cite{gopinath2023comprehensive}. To address this limitation, numerous machine learning–based approaches have been proposed using static, dynamic, and hybrid features \cite{aboaoja2022malwareissues,gorment2023machine}. Classical classifiers, including Logistic Regression, Random Forests, Support Vector Machines, and Gradient Boosting, remain widely used due to their strong performance and interpretability \cite{batouche2021comprehensive,gao2022malwarelightGBM,azeez2021windowspe,gupta2020improving}, while more recent studies explore deep learning architectures to learn representations from raw binaries or execution traces \cite{rahali2020didroid,musikawan2022enhanced,hao2022eii,lu2022self,robinette2024case}.

Although these approaches often achieve high detection accuracy, they typically operate on extremely high-dimensional feature spaces and increasingly complex models. As a result, scalability and interpretability remain persistent challenges, particularly in large-scale deployments \cite{kumar2022scalable,kumar2023scalable,dolejvs2022interpretability,saqib2024comprehensive}. Moreover, many studies emphasize predictive performance without examining whether the underlying feature representations are stable or reproducible across different data partitions \cite{khaire2022stability,yang2024scrr}, limiting their practical reliability.

\subsection{Feature Selection for Malware Detection}

Feature selection plays a critical role in reducing dimensionality and improving generalization in malware detection systems \cite{keyvanpour2023android,dabas2023effective,alomari2023malware}. Existing approaches span filter-based methods, wrapper-based strategies, and embedded techniques based on regularization or tree ensembles \cite{theng2024feature,chandrashekar2014survey}. In particular, feature importance measures derived from Random Forests and Gradient Boosting models are widely adopted due to their ability to capture non-linear feature interactions \cite{pathak2025machine,upadhyay2020gradient,iranzad2025review,adler2022feature}.

However, most existing methods compute global feature importance over the entire dataset, making them sensitive to sampling bias, feature correlation, and class imbalance \cite{rajbahadur2021impact}. Consequently, the selected feature subsets may vary substantially across runs and lack stability, especially in large-scale and heterogeneous datasets. Despite these limitations, locality-aware or chunk-wise feature selection strategies remain relatively underexplored in malware detection, motivating the need for more robust alternatives.

\subsection{Stability and Redundancy in Feature Selection}

Prior work in general machine learning has emphasized the importance of feature stability and redundancy reduction in high-dimensional settings \cite{wang2021feature}. Stability-aware feature selection methods aim to produce consistent feature subsets under data perturbations, while redundancy analysis seeks to eliminate correlated or duplicate features. In malware detection, however, such analyses are rarely performed explicitly. Many studies report feature importance rankings without quantifying stability across data splits or measuring inter-feature correlation \cite{li2020new}. This lack of stability and redundancy evaluation represents a key methodological gap, particularly for security datasets where features are often derived from overlapping rules or heuristics \cite{khaire2022stability,buyukkecceci2023comprehensive}. Addressing this gap requires feature selection frameworks that explicitly account for both stability and redundancy at scale.

\subsection{Explainability in Malware Detection}

Explainability has become increasingly important in malware detection to support analyst trust and regulatory compliance \cite{saqib2024comprehensive,najibi2025towards}. Post-hoc explanation techniques such as SHAP and LIME are commonly applied to interpret model predictions by identifying influential features \cite{dwivedi2023explainable}. While these methods improve transparency, they are often applied to complex models trained on full feature sets, which can obscure the interpretation of individual features. Integrating feature selection with explainability offers a more transparent alternative by limiting explanations to a compact and meaningful feature subset. However, this combination has received limited attention in prior malware detection research, particularly in large-scale experimental settings.

In summary, existing malware detection research prioritizes classification performance and model complexity, while issues related to feature selection stability, redundancy, and scalability remain insufficiently addressed. Guided by the principle of stability-aware and locality-aware feature selection, CAFE-GB differs from prior work by aggregating local importance estimates across data chunks to produce stable and non-redundant feature subsets. By combining systematic feature budget selection, redundancy analysis, statistical validation, and explainability, CAFE-GB directly addresses key gaps in large-scale malware feature selection.

\section{Proposed Methodology}
\label{sec:methodology}

\begin{figure*}[!htbp]
    \centering
    \includegraphics[width=\textwidth]{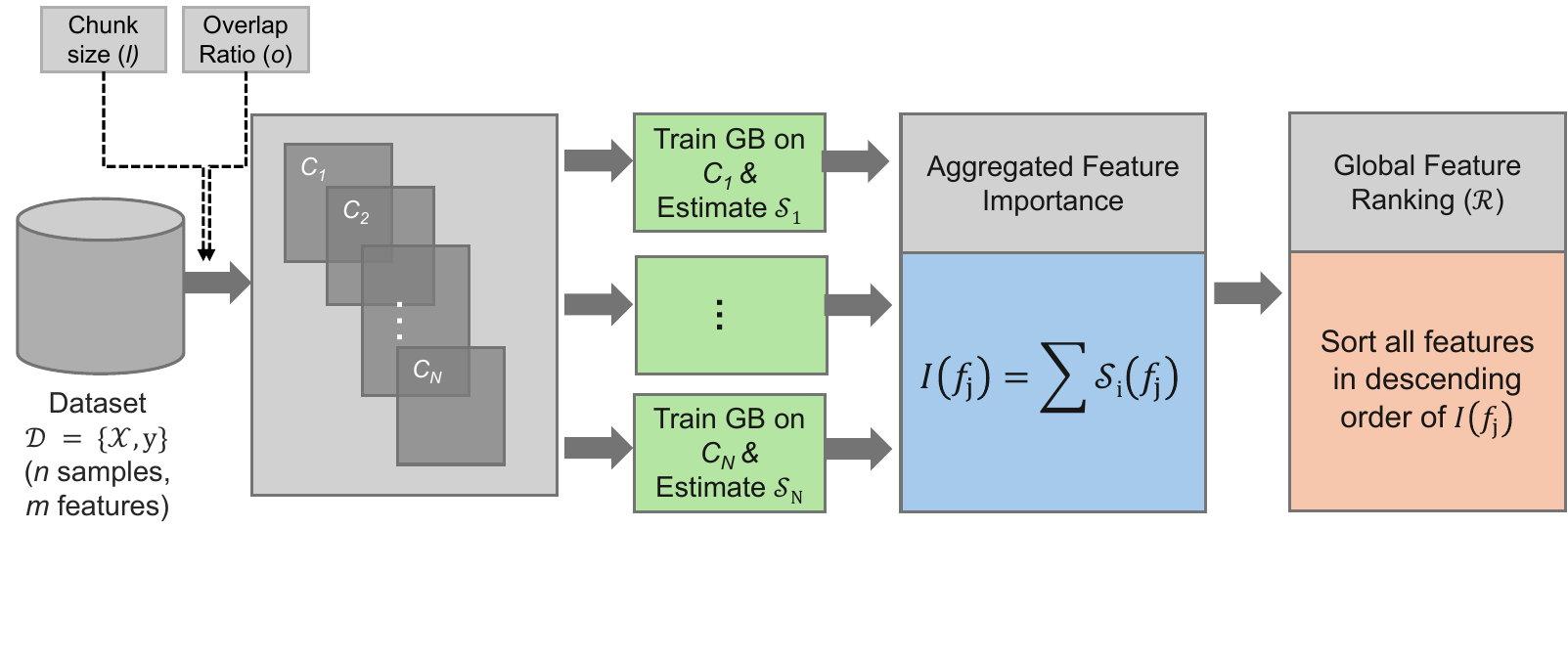}
    \caption{proposed CAFE-GB featture selection framework.}
    \label{fig:cafe-gb-framework}
\end{figure*}

This section presents the proposed CAFE-GB, a chunk-wise feature ranking framework designed to produce stable and robust feature importance estimates for high-dimensional malware datasets. The CAFE-GB leverages gradient boosting, a well-established ML algorithm renowned for its proficiency to handle complex data relationships and provide accurate feature importance metrics. It identifies and prioritizes features that significantly contribute to malware detection, thereby streamlining the feature selection process and enhancing model efficiency. CAFE-GB outputs a global ranking of features and does not perform feature subset size optimization. The architecture of proposed CAFE-GB framework is illustrated in Figure~\ref{fig:cafe-gb-framework} and detailed in Algorithm~\ref{alg:cafe-gb}.
\begin{algorithm}[h]
\caption{CAFE-GB: Chunk-wise Aggregated Feature Estimation using Gradient Boosting}
\label{alg:cafe-gb}

\KwIn{
Training dataset $\mathcal{D} = \{X, y\}$ with $n$ samples and $m$ features; \\
Chunk size $p$; overlap ratio $o$
}

\KwOut{
Global ranked feature list $\mathcal{R}$
}

\BlankLine

\textbf{Step 1: Chunk-wise Data Partitioning} \\
Partition $\mathcal{D}$ into $N$ overlapping chunks 
$\{\mathcal{C}_1, \mathcal{C}_2, \dots, \mathcal{C}_N\}$ using a deterministic sliding window, 
where each chunk contains $p$ samples and consecutive chunks overlap by a ratio $o$.

\BlankLine

\For{$i = 1$ to $N$}{
    \textbf{Step 2: Local Feature Importance Estimation} \\
    Train a gradient boosting classifier on chunk $\mathcal{C}_i$ \;
    Compute local feature importance scores 
    $\mathcal{S}_i = \{f^{i}_1, f^{i}_2, \dots, f^{i}_m\}$ \;
}

\BlankLine

\textbf{Step 3: Aggregated Feature Importance} \\
For each feature $f_j$, compute its aggregated importance score:
\[
I(f_j) = \sum_{i=1}^{N} S_i(f_j)
\]

\BlankLine

\textbf{Step 4: Global Feature Ranking} \\
Sort all features in descending order of $I(f_j)$ to obtain the global ranking $\mathcal{R}$.

\BlankLine

\Return $\mathcal{R}$

\end{algorithm}

\subsection{Chunk-wise Data Partitioning}
The first step involves partitioning the dataset into manageable, overlapping chunks to facilitate comprehensive feature selection. The dataset \(\mathcal{D}=\{X,y\}\) comprises \(n\) instances, where $X \in \mathbb{R}^{n \times m}$ represents the feature matrix with $m$ features, and $y \in \mathbb{R}^{n}$ represents the target labels (malware or benign files). The data partitioning process generates \(N\) overlapping chunks, denoted as \[\mathcal{C}_1, \mathcal{C}_2, \dots, \mathcal{C}_N\] with the following characteristics:
\begin{itemize}
    \item Each chunk $\mathcal{C}_i \in \mathbb{R}^{l \times m}$ contains a fixed number of 
$l$ samples, which is defined based on memory and computational constraints.
    \item Consecutive chunks overlap by a ratio $o$, ensuring that each chunk shares a subset of samples with its predecessor and successor.
\end{itemize}

The overlapping of samples in between chunks ensures that each feature has the opportunity to influence multiple segments of the analysis, thereby enhancing the robustness and reliability of feature selection.

\subsection{Local Feature Importance Estimation}
Once the dataset has been partitioned into overlapping chunks, the next step involves training gradient-boosting classifier to each segment to evaluate feature importance. Gradient boosting is a resilient ensemble learning technique that builds a series of decision trees to predict the target variable\cite{upadhyay2020gradient}. It is known for its capacity to model complex data relationships and provide accurate feature importance metrics. For each chunk $\mathcal{C}_i$, the gradient-boosting classifier is trained using the features $X_i$ and target labels $y_i$ specific to that segment. The model calculates the importance scores for each feature, which gauge how much each feature influences the classifier's predictive accuracy within its respective chunk.

For each chunk $\mathcal{C}_i$, by applying gradient boosting, we estimate the feature importance scores \(\mathcal{S}_i = \{f^{i}_1, f^{i}_2, \dots, f^{i}_m\}\), where \(f^{i}_j\) represents the importance score of the feature \(j\) in chunk $i$. By analyzing each chunk independently, we obtain localized importance scores that capture the relevance of features within specific data segments. This localized approach ensures that feature selection is sensitive to variations across the dataset.

\subsection{Aggregated Feature Importance and Ranking}

To obtain a global importance estimate, CAFE-GB aggregates local importance scores across all chunks.
For each feature $f_j$, the aggregated importance score is computed as:
\[
I(f_j) = \sum_{i=1}^{N} \mathcal{S}_i(f_j)
\]
where \(I(f_j)\) denotes the aggregated importance score for feature \(f_j\), and \(\mathcal{S}_i(f_j)\) is the importance score of feature \(f_j\) in chunk \(\mathcal{C}_i\).
This aggregation emphasizes features that are consistently informative across multiple data segments, while suppressing features that are only locally or spuriously important. The aggregated importance scores are then sorted in descending order to produce the global  feature ranking $\mathcal{R}$.

\section{Datasets and Experimental Setup}
\label{sec:experimental-setup}

\subsection{Datasets}
\label{sec:datasets}

We evaluate the proposed CAFE-GB feature selection framework on two large-scale malware detection datasets: BODMAS\cite{yand2021bodmas}, and CIC-AndMal2020\cite{rahali2020didroid,keyes2021entroplyzer}. These datasets differ in terms of platform, feature dimensionality, data sources, and malware characteristics providing a comprehensive testbed for assessing the scalability, robustness, and effectiveness of CAFE-GB.

\subsubsection{BODMAS}
BODMAS is a large-scale malware dataset containing static features extracted from Windows Portable Executable (PE) files. The dataset consists 77,142 benign and 57,293 malware instances, totaling 134,435 samples, captured during 2019-2020.  Each sample is represented by 2,381 static features derived from byte-level statistics, structural attributes, and metadata. Due to its size and dimensionality, BODMAS is well-suited for evaluating the scalability and effectiveness of feature selection methods in realistic malware detection scenarios.
\subsubsection{CIC-AndMal2020}
CIC-AndMal2020 is a comprehensive Android malware dataset that includes both static and dynamic features extracted from mobile applications. It commbrises a total of $\sim$400k Android applications, evenly split into $\sim$200k malware and  $\sim$200k benign instances. The static features consists of strings, API calls, intents, and permissions, while the dynamic features include system calls, network traffic, and file operations. We consoder static features only for our experiments, where each instance is represented by 9,503 features, making CIC-AndMal2020 one of the largest and most diverse Android malware datasets available. The dataset covers a wide range of malware families and benign applications, making it a challenging benchmark for malware detection. Its heterogeneous feature composition enables assessment of feature selection stability across different malware domains.

\subsection{Experimental Setup and Parameter Configuration}
For each dataset, the training and test splits are created using an 80-20 stratified split to maintain class distribution random seed 42 for reproducibility. All features are standardized using z-score normalization based on training data statistics. Labels are dataset-specific and explicitly configured to avoid ambiguity across datasets. No feature engineering or dimensionality reduction is applied prior to CAFE-GB so that feature selection performance reflects the original high-dimensional input space. For brevity we renamed the features on BODMAS and CIC-AndMal2020 datasets as $F1, F2, ..., F2381$ and $f1, f2, ..., f9503$ respectively.
In our experiments, we set chunk size $l = 15000$ and overlap ratio $o= 0.1$ for all datasets. Using a fixed chunk size ensures consistent local modeling capacity across datasets of different scales and avoids overly coarse partitions in large datasets. This configuration balances computational efficiency with sufficient data representation within each chunk. Gradient Boosting models are trained using the LightGBM implementation with default hyperparameters, which have been shown to perform well across various malware detection tasks \cite{gao2022malwarelightGBM}. Feature importance is computed using the built-in feature importance metric based on gain.

For each dataset, experiments are repeated across five random seeds, and results are reported as mean ± standard deviation. Feature selection is performed independently of classification, followed by downstream model training using a fixed feature budget determined through a dedicated stability analysis. Classification performance is evaluated using Logistic Regression (LR), Random Forest (RF), XGBoost, and LightGBM classifiers. Baseline results correspond to training the same classifiers on the full feature set without feature selection.

\section{Results and Analysis}
\label{sec:results}
This section presents a comprehensive evaluation of the proposed CAFE-GB feature selection framework. The analysis is organized to progressively assess feature selection stability, detection performance, statistical reliability, interpretability, and computational efficiency. All experiments are conducted using identical data splits, classifiers, and random seeds to ensure fair and reproducible comparison.

\subsection{Feature Budget Selection and Stability Analysis}
\label{sec:feature_budget}

\begin{figure*}[!ht]
    \centering
    \includegraphics[width=\textwidth]{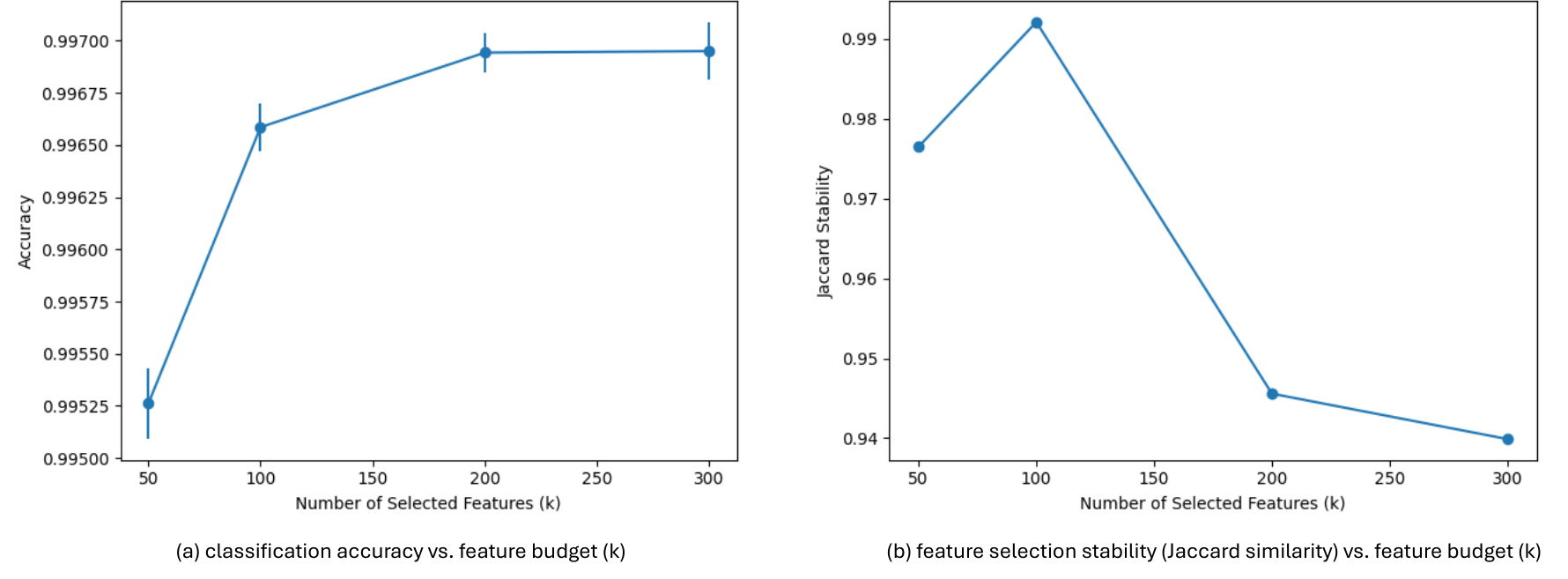}
    \caption{Impact of feature budget on BODMAS}
    \label{fig:bodmas_budget}
\end{figure*}

\begin{figure*}[!ht]
    \centering
    \includegraphics[width=\textwidth]{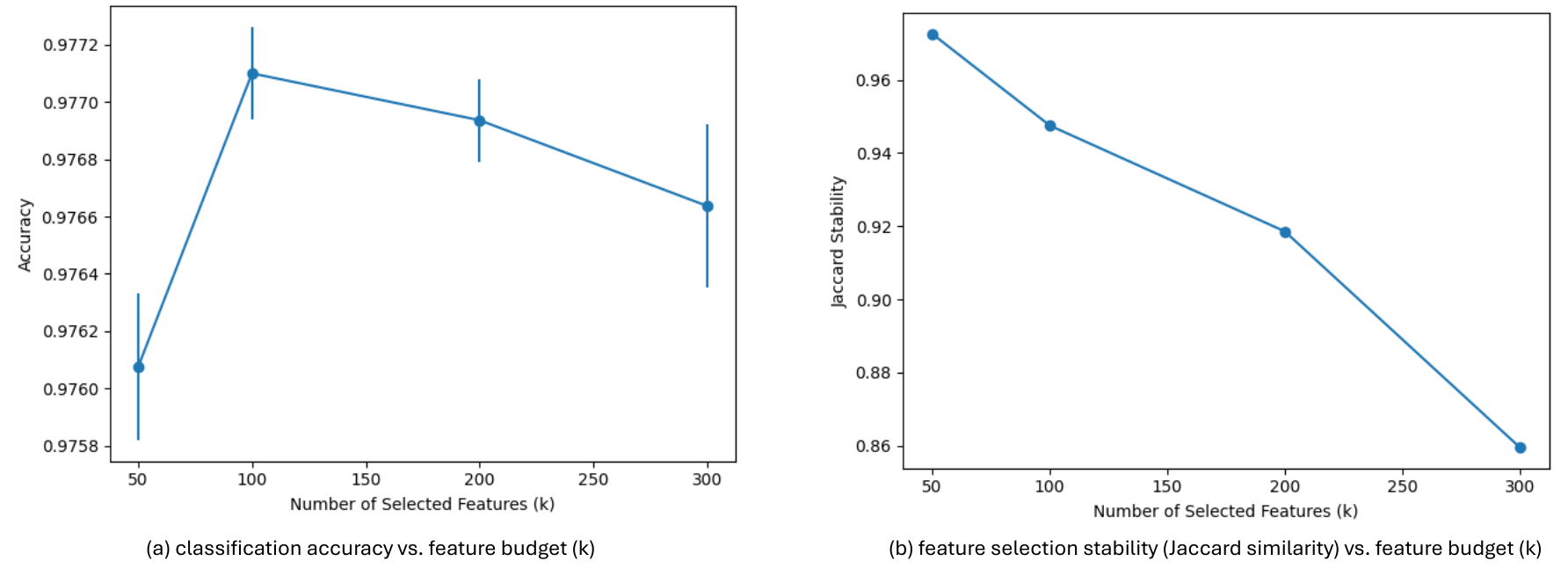}
    \caption{Impact of feature budget on ANDMAL2020}
    \label{fig:andmal_budget}
\end{figure*}

To determine an appropriate feature budget, CAFE-GB is evaluated using feature subset sizes 
\(k \in \{50, 100, 200, 300\}\) on the BODMAS and AndMal2020 datasets. The feature subset sizes were selected to systematically evaluate the trade-off between feature compactness, detection performance, and feature selection stability in high-dimensional malware datasets. These values span small, moderate, and relatively large feature budgets, enabling controlled analysis across different operating regimes.
Table~\ref{tab:feature_budget_analysis} summarizes the average classification accuracy, standard deviation across random seeds, and Jaccard stability of the selected feature subsets. 
Figures~\ref{fig:bodmas_budget} and~\ref{fig:andmal_budget} illustrate the impact of the feature budget on classification accuracy and feature selection stability for each dataset.

\begin{table*}[!ht]
\centering
\small
\caption{Feature Budget Selection and Stability Analysis using CAFE-GB on BODMAS and ANDMAL2020}
\label{tab:feature_budget_analysis}
\begin{tabular}{c|ccc|ccc}
\hline
\multirow{2}{*}{$k$} 
& \multicolumn{3}{c|}{\textbf{BODMAS}} 
& \multicolumn{3}{c}{\textbf{ANDMAL2020}} \\
\cline{2-7}
& Accuracy (Mean) & Accuracy (Std) & Jaccard Stability 
& Accuracy (Mean) & Accuracy (Std) & Jaccard Stability \\
\hline
50  & 0.99526 & 0.00017 & 0.97647 & 0.97608 & 0.00026 & 0.97255 \\
100 & 0.99659 & 0.00011 & 0.99208 & 0.97710 & 0.00016 & 0.94750 \\
200 & 0.99694 & 0.00009 & 0.94557 & 0.97694 & 0.00014 & 0.91855 \\
300 & 0.99695 & 0.00014 & 0.93988 & 0.97664 & 0.00029 & 0.85948 \\
\hline
\end{tabular}
\end{table*}

Across both datasets, the accuracy curves (Figures~\ref{fig:bodmas_budget}(a) and~\ref{fig:andmal_budget}(a)) show a clear improvement when increasing the feature budget from \(k=50\) to \(k=100\). 
Beyond \(k=100\), performance saturates, with only marginal gains or slight degradation observed at \(k=200\) and \(k=300\), indicating diminishing returns from additional features.The stability trends, shown in Figures~\ref{fig:bodmas_budget}(b) and~\ref{fig:andmal_budget}(b), further highlight this trade-off. 
Feature subset stability, measured using Jaccard similarity, peaks at \(k=100\) for both datasets and decreases steadily as the feature budget increases. This decline suggests that larger feature budgets introduce redundant or less informative features, increasing sensitivity to random initialization.

Overall, a feature budget of \(k=100\) provides the most favorable balance between detection performance, robustness, and feature selection stability across both datasets. 
Accordingly, this fixed feature budget is adopted for all subsequent experiments, demonstrating that CAFE-GB supports principled and stability-aware feature budget selection without arbitrary or dataset-specific tuning.


\subsection{Inter-Feature Correlation and Redundancy Analysis}

To evaluate the compactness and redundancy of the feature subsets selected by CAFE-GB, we analyze pairwise Pearson correlations among the top-$k$ selected features ($k=100$). Table~\ref{tab:correlation} reports the mean absolute correlation, maximum absolute correlation, and the proportion of strongly correlated feature pairs, defined as those with $|\rho| > 0.8$. We define strong inter-feature correlation as $|\rho| > 0.8$, a conservative threshold commonly adopted in malware feature selection studies to identify severe redundancy\cite{kong2023malfsm}. Across both datasets, CAFE-GB yields low average inter-feature correlation, with mean $|\rho|$ values of 0.0734 for BODMAS and 0.1788 for AndMal2020. These low mean correlations indicate that, on average, the selected features capture complementary information rather than redundant signals, despite the high dimensionality and sparsity typical of malware feature spaces.

Although the maximum absolute correlation reaches 0.9409 for BODMAS and 1.0 for AndMal2020, such cases are rare. Specifically, only 0.06\% of feature pairs in BODMAS and 0.3\% in AndMal2020 exhibit strong correlation ($|\rho| > 0.8$). This demonstrates that highly correlated feature pairs constitute a negligible fraction of the selected subsets and do not dominate the overall correlation structure. The presence of isolated near-perfect correlations is expected in malware datasets containing binary or rule-derived indicators, where certain features may exhibit deterministic or near-deterministic relationships. Importantly, CAFE-GB does not systematically favor such redundant features; instead, it retains a small number of correlated features only when they are consistently informative across data chunks.

\begin{figure*}[!ht]
    \centering
    \includegraphics[width=\textwidth]{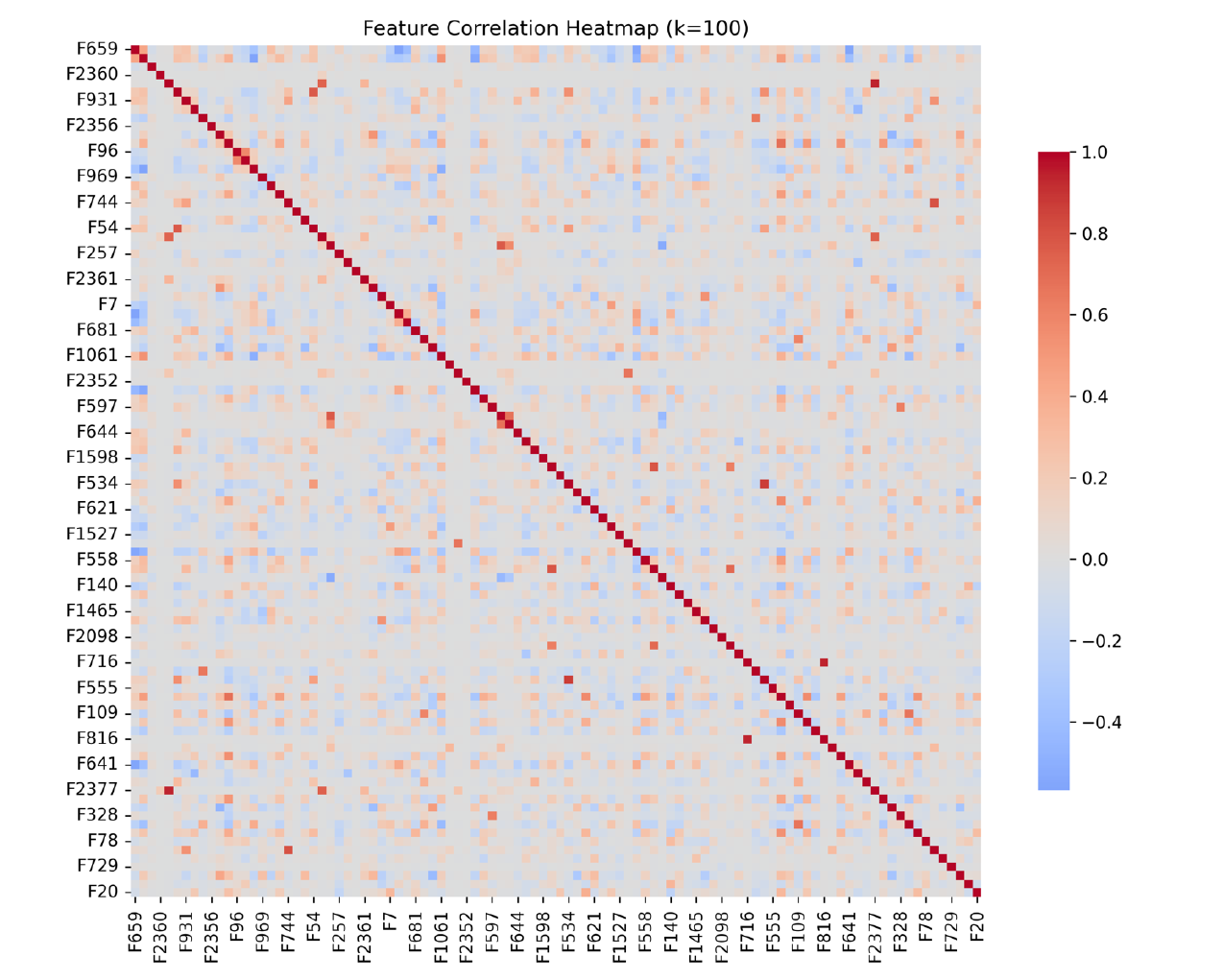}
    \caption{Correlation heatmap for BODMAS dataset with k=100.}
    \label{fig:corr_heatmap_bodmas_k100}
\end{figure*}

Figure~\ref{fig:corr_heatmap_bodmas_k100} presents a representative correlation heatmap for BODMAS ($k=100$). The absence of large contiguous blocks of highly correlated features further confirms that CAFE-GB avoids concentrating importance on tightly coupled feature groups. Similar correlation patterns are observed for AndMal2020 and are omitted for brevity.
Overall, these results indicate that aggregating feature importance scores across data chunks in CAFE-GB effectively suppresses widespread feature redundancy while preserving a compact and informative feature representation suitable for high-dimensional malware detection tasks.

\begin{table*}[!ht]
\centering
\caption{Inter-feature correlation and redundancy analysis for CAFE-GB-selected features ($k=100$).}
\label{tab:correlation}
\begin{tabular}{lcccc}
\hline
\textbf{Dataset} & \textbf{$k$} & \textbf{Mean $|\rho|$} & \textbf{Max $|\rho|$} & \textbf{$|\rho| > 0.8$ (\%)} \\
\hline
BODMAS      & 100 & 0.0734 & 0.9409 & 0.06 \\
AndMal2020  & 100 & 0.1788 & 1.0000 & 0.30 \\
\hline
\end{tabular}
\end{table*}

\subsection{Classification Performance Evaluation}
This section evaluates the classification performance of models trained using features selected by CAFE-GB ($k=100$) compared to baseline models trained on the full feature set. We assess performance across multiple classifiers, including Logistic Regression, Random Forest, XGBoost, and LightGBM, using standard metrics: Accuracy, F1-score, Matthews Correlation Coefficient (MCC), ROC-AUC, and PR-AUC. Results are reported as mean $\pm$ standard deviation over five random seeds to capture both predictive performance and stability.

\subsubsection{Comparison with Baseline (All Features)}
On the BODMAS dataset, CAFE-GB achieves performance that is largely comparable to the baseline across all classifiers, despite operating on a substantially reduced feature set. For tree-based models (LightGBM, Random Forest, and XGBoost), accuracy, F1-score, ROC-AUC, and PR-AUC remain near saturation, indicating that CAFE-GB preserves the discriminative information contained in the full feature space. Notably, CAFE-GB improves the Matthews Correlation Coefficient (MCC) for Random Forest from 0.9875 to 0.9900, suggesting enhanced robustness in capturing both positive and negative class associations under severe class imbalance. For Logistic Regression on BODMAS, accuracy remains stable, while CAFE-GB yields a clear improvement in MCC (0.5820 vs.\ 0.5587). This gain indicates more balanced predictions, even though reductions in ROC-AUC and PR-AUC are observed due to the linear model's limited capacity to exploit higher-order feature interactions.

On the AndMal2020 dataset, CAFE-GB again delivers performance closely matching the baseline for all classifiers. Differences in accuracy and F1-score remain below 0.05\%, while MCC is preserved or slightly improved, particularly for Random Forest (0.9679 vs.\ 0.9672). Importantly, PR-AUC remains stable across both configurations, confirming that CAFE-GB does not compromise minority-class detection despite aggressive feature reduction. Across both datasets, ROC-AUC values remain effectively unchanged, demonstrating that CAFE-GB preserves ranking ability while substantially reducing feature dimensionality. These results confirm that CAFE-GB enhances feature efficiency without degrading predictive performance and, in several cases, improves correlation-based metrics that are especially informative under class imbalance.
\subsubsection{Performance Across Classifiers}

The effectiveness of CAFE-GB varies systematically across classifier families. Tree-based classifiers, including Random Forest, XGBoost, and LightGBM, benefit most from CAFE-GB-selected features, exhibiting high performance and low variance across runs. This behavior is consistent with CAFE-GB’s ability to capture non-linear feature interactions that align naturally with ensemble learning mechanisms. Logistic Regression, despite its linear nature, also demonstrates consistent improvements in MCC across both datasets. This indicates that CAFE-GB produces feature subsets with improved linear separability, enhancing generalization even for simpler models. Overall, the near-equivalence in accuracy and AUC metrics between the baseline and CAFE-GB across all classifiers suggests that performance gains are attributable to improved feature selection rather than increased model complexity. These findings demonstrate that CAFE-GB yields compact, classifier-agnostic feature subsets that maintain or enhance classification performance across diverse learning paradigms.

\begin{table*}[!ht]
\centering
\small
\caption{Classification performance comparison between baseline (all features) and CAFE-GB ($k=100$) on BODMAS and ANDMAL2020 datasets. Results are reported as mean $\pm$ standard deviation.}
\label{tab:classification_bodmas_andmal2020}
\begin{tabular}{lllcccccc}
\toprule
Dataset & Method & Classifier & Accuracy & F1 & MCC & ROC\_AUC & PR\_AUC \\
\midrule
\multirow{8}{*}{BODMAS}
& Baseline & LGBM & 0.9977 $\pm$ 0.0001 & 0.9973 $\pm$ 0.0001 & 0.9953 $\pm$ 0.0001 & 0.9999 $\pm$ 0.0 & 0.9999 $\pm$ 0.0 \\
& Baseline & LR   & 0.7716 $\pm$ 0.0    & 0.7608 $\pm$ 0.0    & 0.5587 $\pm$ 0.0    & 0.8668 $\pm$ 0.0 & 0.8037 $\pm$ 0.0 \\
& Baseline & RF   & 0.9939 $\pm$ 0.0001 & 0.9928 $\pm$ 0.0001 & 0.9875 $\pm$ 0.0002 & 0.9997 $\pm$ 0.0 & 0.9996 $\pm$ 0.0 \\
& Baseline & XGB  & 0.9974 $\pm$ 0.0002 & 0.9970 $\pm$ 0.0002 & 0.9948 $\pm$ 0.0003 & 0.9999 $\pm$ 0.0 & 0.9998 $\pm$ 0.0 \\
& CAFE-GB & LGBM & 0.9972 $\pm$ 0.0001 & 0.9967 $\pm$ 0.0001 & 0.9943 $\pm$ 0.0003 & 0.9998 $\pm$ 0.0 & 0.9998 $\pm$ 0.0 \\
& CAFE-GB & LR   & 0.7711 $\pm$ 0.0032 & 0.7736 $\pm$ 0.0039 & 0.5820 $\pm$ 0.0082 & 0.8054 $\pm$ 0.0012 & 0.6672 $\pm$ 0.0028 \\
& CAFE-GB & RF   & 0.9951 $\pm$ 0.0001 & 0.9942 $\pm$ 0.0001 & 0.9900 $\pm$ 0.0002 & 0.9998 $\pm$ 0.0 & 0.9997 $\pm$ 0.0 \\
& CAFE-GB & XGB  & 0.9971 $\pm$ 0.0001 & 0.9966 $\pm$ 0.0001 & 0.9941 $\pm$ 0.0001 & 0.9998 $\pm$ 0.0 & 0.9998 $\pm$ 0.0 \\
\midrule
\multirow{8}{*}{ANDMAL2020}
& Baseline & LGBM & 0.9795 $\pm$ 0.0003 & 0.9797 $\pm$ 0.0003 & 0.9591 $\pm$ 0.0006 & 0.9973 $\pm$ 0.0001 & 0.9969 $\pm$ 0.0002 \\
& Baseline & LR   & 0.9559 $\pm$ 0.0    & 0.9565 $\pm$ 0.0    & 0.9121 $\pm$ 0.0    & 0.9893 $\pm$ 0.0 & 0.9875 $\pm$ 0.0 \\
& Baseline & RF   & 0.9836 $\pm$ 0.0001 & 0.9837 $\pm$ 0.0001 & 0.9672 $\pm$ 0.0002 & 0.9979 $\pm$ 0.0 & 0.9971 $\pm$ 0.0001 \\
& Baseline & XGB  & 0.9784 $\pm$ 0.0002 & 0.9785 $\pm$ 0.0002 & 0.9569 $\pm$ 0.0005 & 0.9971 $\pm$ 0.0001 & 0.9967 $\pm$ 0.0001 \\
& CAFE-GB & LGBM & 0.9791 $\pm$ 0.0002 & 0.9793 $\pm$ 0.0002 & 0.9583 $\pm$ 0.0004 & 0.9973 $\pm$ 0.0 & 0.9969 $\pm$ 0.0 \\
& CAFE-GB & LR   & 0.9554 $\pm$ 0.0003 & 0.9559 $\pm$ 0.0003 & 0.9110 $\pm$ 0.0007 & 0.9884 $\pm$ 0.0004 & 0.9862 $\pm$ 0.0006 \\
& CAFE-GB & RF   & 0.9839 $\pm$ 0.0001 & 0.9840 $\pm$ 0.0001 & 0.9679 $\pm$ 0.0001 & 0.9978 $\pm$ 0.0001 & 0.9969 $\pm$ 0.0001 \\
& CAFE-GB & XGB  & 0.9790 $\pm$ 0.0002 & 0.9791 $\pm$ 0.0002 & 0.9581 $\pm$ 0.0003 & 0.9971 $\pm$ 0.0 & 0.9968 $\pm$ 0.0 \\
\bottomrule
\end{tabular}
\end{table*}

\subsection{Statistical Significance Analysis}

To evaluate whether the performance differences between the baseline (all features) and the proposed CAFE-GB feature selection are statistically meaningful, we employ a paired Wilcoxon signed-rank test across multiple random seeds. This non-parametric test is well suited for paired comparisons without assuming normality. Table~\ref{tab:stat_significance} reports the corresponding $p$-values and 95\% confidence intervals (CI) for key performance metrics using LightGBM as a representative classifier. On the BODMAS dataset, CAFE-GB ($k=100$) achieves performance that is statistically indistinguishable from the baseline using the full feature set ($k=2381$) across all evaluated metrics. Although slight reductions in mean Accuracy, F1-score, MCC, ROC-AUC, and PR-AUC are observed, none of these differences are statistically significant ($p = 0.0625$ for all metrics). The narrow confidence intervals further indicate that performance variations across random seeds are minimal, confirming the stability of CAFE-GB despite a feature reduction of over 95\%.

Similarly, on the AndMal2020 dataset, no statistically significant differences are detected between the baseline ($k=9503$) and CAFE-GB ($k=100$). All evaluated metrics yield $p$-values well above the conventional significance threshold ($p \geq 0.1875$), with ROC-AUC and PR-AUC exhibiting nearly identical mean values and extremely tight confidence intervals. These results demonstrate that CAFE-GB preserves ranking and minority-class detection performance even under aggressive dimensionality reduction. Overall, the statistical analysis confirms that CAFE-GB achieves \emph{performance parity} with the baseline across both datasets while using substantially fewer features. The absence of statistically significant degradation across all metrics indicates that CAFE-GB effectively removes redundant and non-informative features without compromising predictive reliability. This finding reinforces the practical utility of CAFE-GB as a feature-efficient alternative to full-dimensional models.

\begin{table*}[!ht]
\centering
\caption{Statistical Significance Analysis using Paired Wilcoxon Signed-Rank Test}
\label{tab:stat_significance}
\resizebox{\textwidth}{!}{
\begin{tabular}{l l l c c c c c c c}
\hline
\textbf{Dataset} & \textbf{Classifier} & \textbf{Metric} &
\textbf{Baseline $k$} & \textbf{Proposed $k$} &
\textbf{Mean$_{Base}$} & \textbf{Mean$_{Prop}$} &
\textbf{CI$_{95\%}^{L}$} & \textbf{CI$_{95\%}^{U}$} &
\textbf{$p$-value} \\
\hline
\multirow{7}{*}{BODMAS} 
 & LGBM & Accuracy  & 2381 & 100 & 0.9977 & 0.9972 & 0.9971 & 0.9973 & 0.0625 \\
 & LGBM & F1-score  & 2381 & 100 & 0.9973 & 0.9967 & 0.9966 & 0.9969 & 0.0625 \\
 & LGBM & MCC       & 2381 & 100 & 0.9953 & 0.9943 & 0.9941 & 0.9945 & 0.0625 \\
 & LGBM & ROC-AUC   & 2381 & 100 & 0.9999 & 0.9998 & 0.9998 & 0.9998 & 0.0625 \\
 & LGBM & PR-AUC    & 2381 & 100 & 0.9999 & 0.9998 & 0.9998 & 0.9998 & 0.0625 \\
\hline
\multirow{7}{*}{AndMAL2020}
 & LGBM & Accuracy  & 9503 & 100 & 0.9795 & 0.9791 & 0.9790 & 0.9793 & 0.1875 \\
 & LGBM & F1-score  & 9503 & 100 & 0.9797 & 0.9793 & 0.9791 & 0.9794 & 0.1875 \\
 & LGBM & MCC       & 9503 & 100 & 0.9591 & 0.9583 & 0.9580 & 0.9586 & 0.1875 \\
 & LGBM & ROC-AUC   & 9503 & 100 & 0.9973 & 0.9973 & 0.9972 & 0.9973 & 0.3125 \\
 & LGBM & PR-AUC    & 9503 & 100 & 0.9969 & 0.9969 & 0.9969 & 0.9970 & 0.6250 \\
\hline
\end{tabular}
}
\end{table*}

\subsection{Explainability Analysis}

\begin{figure*}[!ht]
    \centering
    \includegraphics[width=\textwidth]{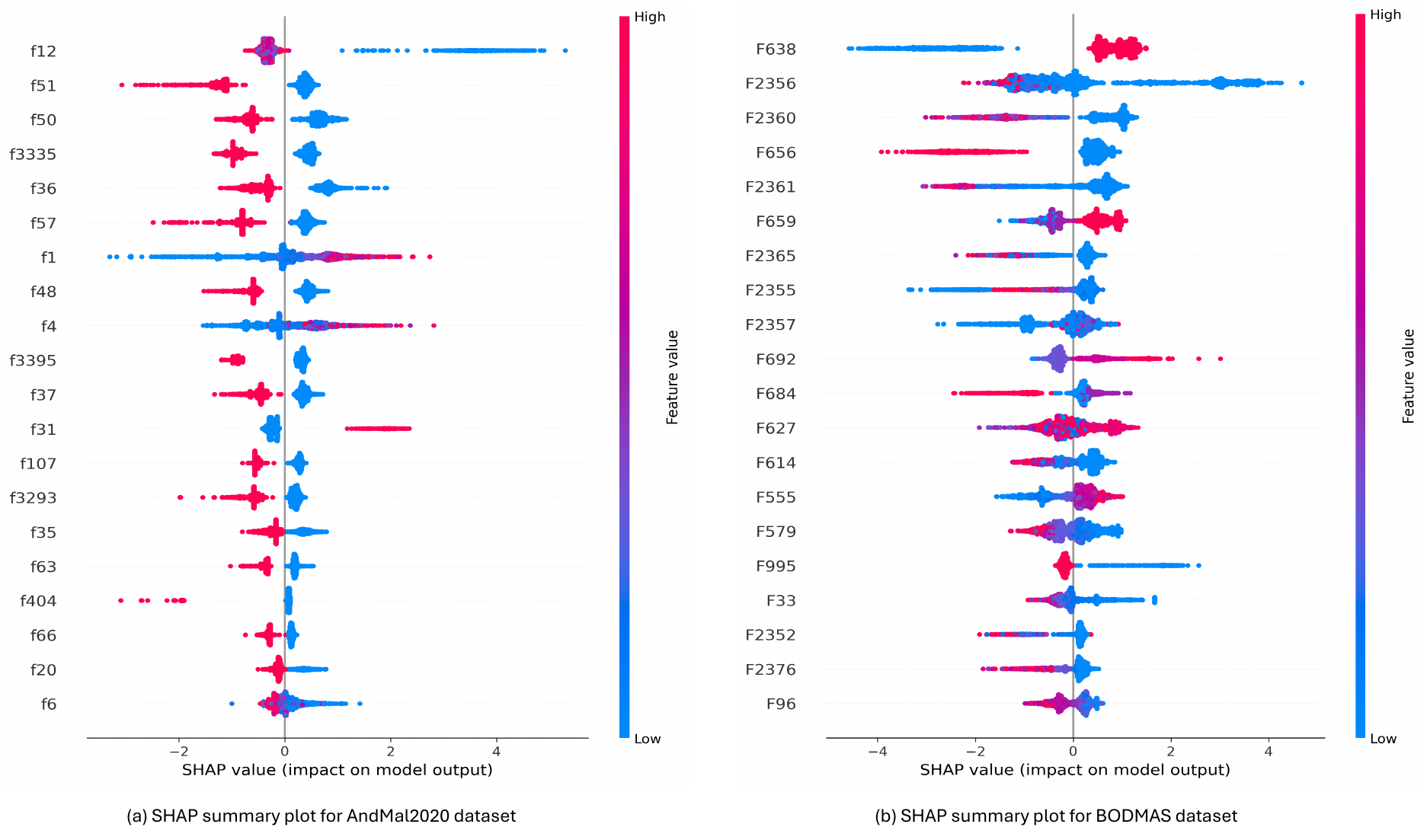}
    \caption{SHAP summary plots for the LightGBM model on AndMal2020 and BODMAS datasets. The plot shows the impact of each feature on the model's output, with features ordered by their importance. Positive SHAP values indicate a higher likelihood of the positive class (malware), while negative values indicate a lower likelihood.}
    \label{fig:shap-summary-all}
\end{figure*}

To assess the interpretability of CAFE-GB-selected features, we apply SHAP-based post-hoc explainability to models trained using the reduced feature sets. The analysis is conducted using the LightGBM classifier with a fixed random seed (42) to provide a controlled and reproducible explanation setting. Figures~\ref{fig:shap-summary-all}(a) and~\ref{fig:shap-summary-all}(b) present SHAP summary plots for the AndMal2020 and BODMAS datasets, respectively, restricted to the top 20 most influential features. Across both datasets, the SHAP analysis reveals a strong concentration of predictive importance in a relatively small subset of features, indicating that CAFE-GB effectively removes redundant and weakly informative attributes. This concentration supports the objective of producing compact and interpretable feature representations without sacrificing discriminative capability.

Although the specific features differ across datasets, a consistent pattern emerges at the semantic level: in both cases, the dominant SHAP features correspond to behaviorally meaningful malware characteristics rather than spurious correlations. The clear separation between positively and negatively contributing features further enhances transparency by allowing direct attribution of model predictions to specific behavioral indicators. Overall, this analysis demonstrates that CAFE-GB-selected features support interpretable decision-making across heterogeneous malware domains, improving model transparency while maintaining high predictive performance.

\subsection{Computational Efficiency and Scalability}

We evaluate the computational efficiency and scalability of CAFE-GB by measuring end-to-end runtime and peak memory consumption across multiple random seeds on both the BODMAS and AndMal2020 datasets. Table~X summarizes the observed execution costs for the feature selection stage (CAFE-GB) and the downstream classification stage.
\begin{table*}[!ht]
\centering
\small
\caption{Computational runtime and memory consumption of CAFE-GB feature selection and downstream classification across datasets and random seeds.}
\begin{tabular}{l l c r r}
\toprule
\textbf{Stage} & \textbf{Dataset} & \textbf{Seed} &
\textbf{Runtime (s)} & \textbf{Memory (MB)} \\
\midrule
\multirow{5}{*}{CAFE-GB}
 & BODMAS       & 42 & 3024.01 & 2423.18 \\
 & BODMAS       & 52 & 3168.76 & 3079.06 \\
 & BODMAS       & 62 & 3037.42 & 3245.47 \\
 & BODMAS       & 72 & 3016.82 & 3351.66 \\
 & BODMAS       & 82 & 3141.07 & 3464.29 \\
\midrule
\multirow{5}{*}{CAFE-GB}
 & AndMal2020   & 42 & 2150.21 & 29338.48 \\
 & AndMal2020   & 52 & 1198.94 & 32643.38 \\
 & AndMal2020   & 62 & 1203.20 & 33758.43 \\
 & AndMal2020   & 72 & 1182.41 & 34691.55 \\
 & AndMal2020   & 82 & 1177.84 & 35184.38 \\
\midrule
\multirow{5}{*}{Classification}
 & BODMAS       & 42 & 899.50  & 4352.75 \\
 & BODMAS       & 52 & 1052.90 & 7190.88 \\
 & BODMAS       & 62 & 1368.46 & 7418.02 \\
 & BODMAS       & 72 & 1294.46 & 7694.16 \\
 & BODMAS       & 82 & 814.44  & 7768.50 \\
\midrule
\multirow{5}{*}{Classification}
 & AndMal2020   & 42 & 3591.24 & 43094.38 \\
 & AndMal2020   & 52 & 3684.45 & 73851.78 \\
 & AndMal2020   & 62 & 3627.75 & 75950.77 \\
 & AndMal2020   & 72 & 3622.32 & 81380.03 \\
 & AndMal2020   & 82 & 3981.00 & 82130.06 \\
\bottomrule
\end{tabular}
\end{table*}

\paragraph{Runtime Analysis}
On the BODMAS dataset, CAFE-GB exhibits stable execution times across seeds, with runtimes ranging from approximately 3017 to 3169 seconds. This consistency indicates that the proposed feature selection process scales reliably with respect to random initialization. In comparison, the classification stage requires substantially less time, varying between 814 and 1368 seconds, reflecting the reduced computational burden once a compact feature subset has been selected. For the AndMal2020 dataset, CAFE-GB incurs higher absolute runtime due to the larger feature space and increased data complexity. Nevertheless, execution times remain within a bounded range (approximately 1178–2150 seconds), with the majority of runs completing close to 1200 seconds. In contrast, the classification stage is significantly more time-consuming on AndMal2020, requiring approximately 3600–4000 seconds across seeds. This observation highlights that the dominant computational cost in large-scale settings arises from model training rather than from CAFE-GB feature selection.

\paragraph{Memory Consumption}
Memory usage follows a similar dataset-dependent trend. On BODMAS, CAFE-GB requires between 2.4\,GB and 3.5\,GB of memory, which is consistently lower than the classification stage, where memory usage increases up to approximately 7.8\,GB. This demonstrates that CAFE-GB does not introduce excessive memory overhead beyond standard model training. On AndMal2020, memory requirements are higher due to the high-dimensional Android feature space. CAFE-GB consumes between 29\,GB and 35\,GB of memory across seeds, while the classification stage requires substantially more memory, reaching up to 82\,GB. Importantly, CAFE-GB consistently operates within a significantly smaller memory footprint than downstream classification, despite performing feature interaction modeling and global importance aggregation.

\paragraph{Scalability Implications}
Overall, these results indicate that CAFE-GB scales effectively to large, high-dimensional malware datasets. While feature selection introduces a one-time computational cost, this cost is amortized by reduced training complexity and memory usage in subsequent classification stages. The consistent runtime and memory profiles across random seeds further demonstrate that CAFE-GB provides predictable and scalable performance, making it suitable for practical deployment in large-scale malware analysis pipelines.

\section{Discussion}
\label{sec:discussion}

The experimental results demonstrate that CAFE-GB effectively addresses feature redundancy and selection instability in large-scale malware datasets while maintaining scalability and interpretability. Across both evaluated datasets, classifiers trained on CAFE-GB-selected features achieve performance parity with baseline models trained on full feature sets, despite reducing feature dimensionality by more than 95\%. This finding highlights that stable and compact feature selection can preserve detection reliability without increasing model complexity, particularly under severe class imbalance where metrics such as MCC and PR-AUC are most informative. The correlation and redundancy analysis further clarifies the behavior of CAFE-GB. Despite operating on high-dimensional feature spaces, the selected feature subsets exhibit low average inter-feature correlation, indicating that the chunk-wise aggregation strategy promotes diversity among selected features rather than favoring tightly coupled feature groups. This behavior distinguishes CAFE-GB from conventional global feature selection approaches, which may inadvertently amplify correlated or dominant features. At the same time, the reliance on tree-based importance measures introduces a methodological limitation, as such measures can be influenced by feature variance or marginal effects. While aggregation improves stability, integrating alternative attribution mechanisms could further enhance robustness in future extensions.

The systematic k-selection and stability analysis shows that a fixed feature budget can achieve a favorable trade-off between robustness, reproducibility, and efficiency across datasets. Selecting a stable feature budget simplifies experimental comparison and supports reproducibility; however, this design choice may not be optimal for all operational settings. Adaptive or task-dependent feature budgets therefore represent a promising direction for extending CAFE-GB to dynamic or resource-constrained environments. Explainability analysis using SHAP reinforces the interpretability of CAFE-GB-selected features by revealing that a small subset of features consistently dominates model predictions. Importantly, these features correspond to semantically meaningful malware characteristics across both datasets. Nevertheless, SHAP is applied post hoc and captures global feature influence rather than evolving feature relevance. Integrating explainability more tightly with the feature selection process remains an open challenge. From a computational perspective, runtime and memory profiling indicate that CAFE-GB incurs an upfront cost due to chunk-wise processing but substantially reduces downstream classification overhead. This trade-off is acceptable in large-scale offline training scenarios and supports the use of CAFE-GB as a preprocessing step. However, the current evaluation does not consider real-time or streaming deployment settings, which would require further optimization and incremental processing strategies. Overall, the results validate CAFE-GB as a practical and effective feature selection framework for large-scale malware detection. 

While CAFE-GB effectively addresses stability, redundancy, and scalability challenges in large-scale malware feature selection, several limitations warrant discussion. First, the framework relies on tree-based importance measures, which may exhibit bias toward features with strong marginal effects; although chunk-wise aggregation improves robustness, alternative or model-agnostic importance estimators could further enhance stability. Second, the evaluation focuses on static and aggregated behavioral features, and the applicability of CAFE-GB to other modalities—such as raw byte sequences, graph-based representations, or streaming traces—remains to be explored. Third, although CAFE-GB is designed to handle heterogeneous data distributions, explicit temporal generalization and concept drift are not addressed in this study. Finally, a fixed feature budget is adopted to ensure reproducibility and fair comparison; adaptive or task-specific budgets may offer additional flexibility in dynamic settings. Addressing these limitations represents promising directions for future research, including adaptive feature budgeting, alternative importance estimation, and tighter integration of explainability within the feature selection process.

\section{Conclusion}
\label{sec:conclusion}
This paper presented CAFE-GB, a chunk-wise aggregated feature selection framework that departs from conventional global importance estimation by explicitly targeting feature selection stability in large-scale malware detection. By aggregating local gradient boosting–based importance estimates across overlapping data chunks, CAFE-GB derives a globally consistent and stable feature ranking without relying on dataset-specific tuning. Experiments on the BODMAS and CIC-AndMal2020 datasets demonstrate that CAFE-GB achieves substantial dimensionality reduction while preserving detection performance across multiple classifiers. A stability-aware feature budget analysis enables a principled selection of a fixed feature subset, and inter-feature correlation analysis confirms that the resulting representations are compact and low in redundancy. Statistical significance testing shows that performance using CAFE-GB-selected features remains statistically comparable to full-feature baselines, despite aggressive feature reduction. SHAP-based analysis further indicates that the selected features capture semantically meaningful malware characteristics. Overall, CAFE-GB introduces a stable, interpretable, and scalable alternative to conventional feature selection methods for large-scale malware detection.

\section*{Conflict of Interest}
The authors have no conflicts of interest to declare.


\section*{Declaration of generative AI and AI-assisted technologies in the manuscript preparation process}
During the preparation of this work the author(s) used ChatGPT, Gemini and Paperpal in order to improve language and readability of the work. After using this tool/service, the author(s) reviewed and edited the content as needed and take(s) full responsibility for the content of the published article.



\bibliographystyle{elsarticle-num-names} 
\bibliography{references}

\end{document}